\documentclass[12pt,thmsa]{article}%
\usepackage{sw20elba}
\usepackage{amsmath}
\usepackage{graphicx}%
\usepackage{amsfonts}%
\usepackage{amssymb}

\begin{document}

\begin{center}
\textbf{Enhancement of Critical Current across Single Grain Boundaries in
YBa}$_{\mathbf{2}}$\textbf{Cu}$_{\mathbf{3}}$\textbf{O}$_{\mathbf{x}}$\textbf{
and Search for Superconductivity in Alkali Metals}
\end{center}

\begin{center}
\bigskip T. Tomita$^{a}$, S. Deemyad$^{a}$, J.J. Hamlin$^{a}$, J.S.
Schilling$^{a}$, V.G. Tissenb, B.W. Veal$^{c}$, L. Chen$^{c}$, H. Claus$^{c}$

$^{a}$\textit{Department of Physics, Washington University, CB 1105, St.
Louis, MO 63130, USA}

$^{b}$\textit{Institute of Solid State Physics, Chernogolovka 142432, Moscow
District, Russia}

$^{c}$\textit{Materials Science Division, Argonne National Labs, 9700 S. Cass
Ave., Argonne, IL 60439, USA}

\bigskip

\textbf{Abstract}

\smallskip
\end{center}

The ability of single grain boundaries in YBa$_{2}$Cu$_{3}$O$_{x}$
bicrystalline rings to carry electrical current is found to be significantly
enhanced under hydrostatic pressure. Nearly hydrostatic pressures are applied
to the alkali metals Li, Na, and K. in a search for superconductivity.
\ Whereas Li becomes superconducting above 20 GPa at temperatures as high as
15 K, no superconductivity was observed above 4 K in Na to 65 GPa nor in K
above 4 K to 43.5 GPa or above 1.5 K to 35 GPa.

\section{Introduction}

The first high pressure experiments on a superconductor were carried out in
1925 by Sizoo and Onnes \cite{sizoo}, 14 years after the discovery of the
first superconductor \cite{holst}. Since then high-pressure investigations
have had an important impact on the field. Following the discovery of a new
superconductor, one of the first experiments is to determine the pressure
dependence of the superconducting transition temperature $T_{c}.$ If the
magnitude of $dT_{c}/dP$ is large, one can have good hope that higher values
of $T_{c}$ at ambient pressure are possible. In addition, a comparison of the
pressure dependences of $T_{c}$ with those of selected normal state properties
provides information regarding the mechanism(s) responsible for the
superconductivity, as recently illustrated for MgB$_{2}$, the high-$T_{c} $
oxides, and the alkali-doped fullerenes \cite{schilling1}.

In this paper we give further examples to illustrate how the high-pressure
technique can provide vital information and uncover new physics for two
completely different classes of superconductors: \ (1) the high-pressure
enhancement of the critical current density $J_{c}$ across a single grain
boundary (GB) in bicrystalline rings of the high-$T_{c}$ oxide YBa$_{2}%
$Cu$_{3}$O$_{x}$ (YBCO), and (2) pressure-induced superconductivity in the
alkali metals. The observation of pressure-induced superconductivity in Li is
at odds with the conventional wisdom that metals become more
free-electron-like under pressure.

\section{Results on YBCO Grain Boundaries}

Shortly after the discovery of the high-$T_{c}$ oxides in 1986 it became clear
that the $J_{c}$-values for polycrystalline materials not only are vastly
inferior to the those obtained in single crystalline samples, but also degrade
sharply in a magnetic field \cite{mannhart1}. Experiments on epitaxially-grown
YBCO thin films demonstrated that this strong degradation in $J_{c}$ resulted
from the presence of grain boundaries in the polycrystals \cite{chaudhari1};
for adjacent grains with parallel $c$-axes, $J_{c}$ was found to decrease
roughly exponentially with increasing misorientation angle $\theta$
\cite{dimos1}. Pinpointing the mechanisms behind this $J_{c}$ degradation and
developing strategies to enhance $J_{c}$ to the values needed for many
applications have been at the focus of intensive research for many years.

Possible candidates for the $J_{c}$ degradation include various types of
imperfections in or at the GB such as site disorder, lattice strains, and
oxygen vacancies \cite{mannhart1,caldwell,chisholm}. High-pressure experiments
may shed some light on these matters by modifying the conditions at the GB in
several different ways: \ (1) reduction of tunneling barrier width, (2)
varying the degree of lattice strain, (3) enhancing the degree of oxygen
ordering in the GB, in analogy with the well studied pressure-induced oxygen
ordering effects in the bulk \cite{fietz1,sadewasser1}. The very existence of
relaxation effects due to oxygen ordering in the GB would signal that the
oxygen sublattice is only partially full and thus capable of accepting further
oxygen doping.

Previous $ac$ susceptibility studies on bulk polycrystalline samples of
YBa$_{2}$Cu$_{4}$O$_{8}$ and Tl$_{2}$CaBa$_{2}$Cu$_{2}$O$_{8}$ indicated that
the bulk $J_{c}$ was enhanced through pressure, but it was not possible to
extract detailed information \cite{diederichs1}. Studies of pressure-dependent
effects for a single grain boundary are clearly needed. Such experiments have
recently become possible with the availability of bicrystalline rings of
YBa$_{2}$Cu$_{3}$O$_{x}$ with varying misorientation angles $\theta$ and
oxygen content $x$ \cite{claus1}.

The He-gas pressure cell (Unipress) suitable for hydrostatic pressure studies
to 1.4 GPa is shown in Fig. 1. The YBCO bicrystalline ring with typical
dimensions 5 mm O.D. x 3 mm I.D. x 1 mm is mounted in the sample holder made
of Vespel which is placed in the 7 mm bore of the BeCu pressure cell. An
alumina cylinder is positioned above the sample holder to reduce the volume of
He in the cell.

Two counterwound pickup coils are used in the $ac$ susceptibility
measurement,\ one positioned around the YBCO ring, the other 2.3 mm below. To
vary the temperature over the range 6 - 300 K the pressure cell is placed in
the sample tube of a closed-cycle refrigerator (Balzers) with the BeCu
capillary tube (3 mm O.D. x 0.3 mm I.D.) exiting out the top and connected to
a three-stage He-gas compressor system (Harwood) to 1.4 GPa with a digital
manganin gauge. Since primary $ac$ field amplitudes as high as 300 G are
required, internal heating effects are avoided by removing the field coil from
the pressure cell and placing it outside the tail piece of the cryostat. To
ensure full field penetration to the sample, the frequency of the $ac$ field
was reduced to 1 Hz. Standard $ac$ susceptibility techniques are used with a
SR830 digital lockin amplifier.

In Fig. 2 (left) the real part of the $ac$ susceptibility $\chi_{ac}$ is
plotted versus temperature for an $ac$ field amplitude of 1 G at both ambient
and 0.6 GPa hydrostatic pressure. The sample studied is a YBa$_{2}$Cu$_{3}%
$O$_{6.9}$ bicrystalline ring with misorientation angle $\theta=30^{\circ}.$
At temperatures below 88 K at ambient pressure, the applied flux is not able
to penetrate into the ring since the induced current density through the two
grain boundaries in the ring lies below the critical value $J_{c}$ . Since
$J_{c}(T)$ decreases monotonically with increasing temperature, as the
temperature is raised, a value $T=87.9$ K is reached where flux begins to
penetrate through the weaker of the two grain boundaries into the interior of
the ring; this leads to a sharp break in both the real and imaginary parts of
$\chi_{ac}$ (see vertical arrow in Fig. 2 (left)) at the kink temperature
$T_{kink}.$ As the temperature is increased further, additional flux
penetrates into the center of the ring, but not through the superconducting
material of the ring itself, until a plateau is reached above 90 K. Above 91.7
K magnetic flux begins to penetrate into the ring material until above 92 K
all superconductivity has been destroyed and the flux distribution is uniform.
These results from $ac$ susceptibility measurements are in excellent agreement
with parallel $dc$ susceptibility measurements using a SQUID magnetometer on
the same YBCO ring.

The shielding current $I$ around the ring is directly proportional to the
field amplitude $H,$ $I=DH,$ where,\ to a good approximation, $D$ is the outer
diameter of the ring \cite{claus1}. In Fig. 2 (left) it is seen that under 0.6
GPa pressure $T_{kink}$ shifts to higher temperatures, i.e. $J_{c}$ at a given
temperature increases. This means that under hydrostatic pressure the ability
of the GB to carry current is enhanced. This result is confirmed by further
studies at magnetic field amplitudes to $\sim$ 23 G which suppress $T_{kink}$
to temperatures below 20 K, as seen in Fig. 2 (right). In this figure the
calculated values for $J_{c}$ are given, where $J_{c}=I_{c}/A$ and $A$ is the
cross-sectional area of the ring. At all measured temperatures $J_{c}$ is seen
to increase rapidly with hydrostatic pressure at the rate +26 \%/GPa which is
much more rapid than that of $T_{c}$ for the bulk material ($\sim$ 0.24 \%/GPa).

We also find the change in $J_{c}$ to be less if the pressure is varied at
temperatures significantly below ambient; time dependences in $J_{c}$ are also
observed. Similar relaxation effects are well known in bulk YBCO where $T_{c}$
is found to depend on the detailed pressure/temperature history of the sample
\cite{fietz1,sadewasser1}. The extreme sensitivity of these relaxation effects
on the oxygen content points to the mobile oxygen ions in the basal plane, the
``chain layer'', as being their source. The bulk relaxation effects are
smallest when the chain layer is fully developed and essentially all chain
sites are filled, i.e. at the stoichiometry YBa$_{2}$Cu$_{3}$O$_{7.0}$. In
analogy, the relaxation effects observed in $J_{c}$ for the GB give evidence
that the GB contains a sizeable number of oxygen vacancies, i.e. there's room
for more oxygen! If a way can be found to further enhance the oxygen
concentration in the GB region, it is likely that higher values of $J_{c}$ can
be attained.

Further research is underway to determine the dependence of $J_{c}$ on
pressure as a function of both temperature and $dc$ magnetic field over a wide
range of oxygen content and misorientation angle $\theta$ \cite{tomita1}.

\section{Results on Alkali Metals}

The nearly free electron picture works best for simple $s,p$-electron metals
such as the alkali metals or noble metals. If a simple metal happens to be
superconducting at ambient pressure (e.g. Pb, In, Sn, Al), $T_{c}$ is
invariably found to \textit{decrease} under hydrostatic pressure. This result
has a simple explanation \cite{schilling1}. According to the well-known
McMillan formula, $T_{c}$ depends exponentially on the electron-phonon
coupling parameter $\lambda=\eta/\kappa,$ where $\eta$ is a purely electronic
term, the Hopfield parameter, and $\kappa\equiv M\left\langle \omega
^{2}\right\rangle $ is a mean ``spring constant'' for the lattice. If
$\lambda$ decreases, so does $T_{c}.$ The reason for the universal decrease in
$T_{c}$ under pressure for simple metals is that $\kappa$ increases much more
rapidly than does $\eta.$ Another way to express this is that $T_{c}$
decreases under pressure due to lattice stiffening. In transition metals the
increase of $\eta$ with pressure is comparable to that of $\kappa,$ so that
$T_{c}$ may increase or decrease.

Since for simple-metal superconductors $T_{c}$ always \textit{decreases} under
pressure, it would appear unlikely that a nonsuperconducting simple metal,
like any of the alkali metals, would ever become superconducting under
pressure. Over 30 years ago, however, Wittig \cite{wittig1} demonstrated that
at pressures above 7 GPa the alkali metal Cs becomes superconducting near 1.5
K. At first glance this result would appear to contradict our above
conclusions, but it does not. At pressures above 7 GPa Cs is no longer a
simple metal! Near these pressures $s-d$ transfer occurs as the bottom of the
5$d$ band drops below the top of the 6$s$ band, thus turning Cs into a
transition metal \cite{ross1}. A similar scenario would be expected for the
next lighter alkali metals Rb and K.

In this connection the alkali metal Li is of particular interest. Its
unoccupied 3$d$ band lies far above the Fermi energy; indeed, Li lies two rows
up in the periodic table from the 3$d$ transition metal series. In this
picture, therefore, no pressure-induced superconductivity would be expected in
Li, except at pressures well above the 100 GPa range. In 1986 Lin and Dunn
\cite{lin1} did report a possible phase transition in Li under
quasihydrostatic pressures above 20 GPa, but the evidence for
superconductivity was not unequivocal.

In 1998 Neaton and Ashcroft \cite{neaton1} made the alarming prediction that
at sufficiently high pressures Li should cease to behave like a simple metal,
but rather should show marked deviations from free-electron-like behavior.
Because of core overlap the bands near $E_{f}$ actually become narrower and
band gaps increase markedly, thus contradicting basic high-pressure wisdom
that under pressure bands broaden and band gaps decrease! This result was
mentioned in an earlier paper by Boettger and Trickey \cite{boettger}. At
pressures sufficient to generate overlap of the Li-ion cores, the
orthogonality requirements of the valence electrons to the core states leads
to a marked enhancement in the pseudopotential and thus in the
electron-lattice coupling. This in turn would be expected to generate
structural phase transitions to lower symmetry structures and possible
superconductivity. Extensive experimental \cite{syassen1,hanfland1} and
theoretical \cite{ackland} investigations of structural phase changes in the
alkali metals have confirmed these general expectations. In 2001 Christensen
and Novikov \cite{christensen} predicted from \textit{ab initio} electronic
structure calculations that for $fcc$ Li $T_{c}$ would increase rapidly with
pressure to values approaching 80 K. These predictions prompted several
experimental groups to search in earnest for superconductivity in Li under pressure.

The first group to confirm a superconducting transition in Li was that of
Shimizu \textit{et al. }\cite{shimizu1} in 2002 in a diamond-anvil cell
experiment to 50 GPa. In their electrical resistivity measurement
superconductivity appeared at 7.5 K for 30 GPa pressure, rising rapidly to
values approaching 20 K; the pressure dependence $T_{c}(P)$ was combined from
four separate experiments and exhibited considerable scatter. A few months
later Struzhkin \textit{et al.} \cite{struzhkin1} measured the $ac$
susceptibility to 40 GPa and the resistivity to 82 GPa, reporting an onset of
superconducting at 10 K for 23 GPa with a rapid rise under pressure to $\sim$
16 K at 35 GPa.

In both these studies no pressure medium was used, the diamond anvils being
allowed to press directly onto the Li sample. This raises the question whether
shear stresses on the Li in such experiments may have played a role in the
reported superconducting transition. Such effects have been well documented.
In the first high-pressure ever carried out on a superconductor, Sizoo and
Onnes \cite{sizoo} reported for both Sn and In that $T_{c}$ decreased under
hydrostatic pressure, but \textit{increased} if uniaxial pressure was applied.
If Li metal is cooled from ambient temperature, it undergoes a martensitic
phase transformation at $T_{pt}\simeq75$ K from the $bcc$ to a $Rh6$
low-temperature phase. Under hydrostatic pressure $T_{pt}$ increases at the
rate +30 K/GPa \cite{smith1}, but under uniaxial pressure $dT_{pt}%
/dP\simeq+1250$ K/GPa \cite{maier1}, a rate 40$\times$ faster! In the present
experiment, in fact, nonhydrostatic shear stresses on the Re gasket resulted
in a sharp increase in Re's superconducting transition temperature to 3.5 K,
thus preventing the detection of superconductivity from the Li sample at lower
temperatures. Further examples of sharply differing hydrostatic versus
uniaxial pressure effects on superconductivity are given in Ref.
\cite{deemyad1}. For this reason we set out to search for superconductivity in
Li using the most hydrostatic pressure medium of all, dense He.

In Fig. 3 (left) we show the coil arrangement we used for $ac$ susceptibility
measurements in our nonmagnetic BeCu diamond-anvil cell. Two identical coils
are shown wound with 60 $\mu m$ Cu wire consisting of an outer primary coil
(130 turns) and an inner secondary coil (180 turns). One coil is placed around
the lower diamond anvil (1/6 carat, 0.5 mm culet) with the compensating coil
directly adjacent. Both coils and lead wires are thermally anchored to an
insulating board with GE varnish. A preindented Re gasket is placed in both
measuring and compensating coils to reduce the background signal in the $ac$
susceptibility. Before applying pressure, liquid He is filled into the gasket
hole to serve as a nearly hydrostatic pressure medium. Further experimental
details are given in Ref. \cite{deemyad1}.

In Fig. 4 we show the superconducting phase diagram for Li metal under nearly
hydrostatic pressures to 67 GPa \cite{deemyad1}; this diagram differs
significantly from the results of the previous
studies\cite{lin1,shimizu1,struzhkin1}, particularly above 30 GPa. In the
first run 25 data points were obtained as a function of both increasing and
decreasing pressure. We find that under nearly hydrostatic pressure Li becomes
superconducting at $\sim$ 5 K for 20 GPa, $T_{c}$ increasing initially rapidly
with pressure to $\sim$ 14 K at 30 GPa. At this pressure a structural phase
transition appears to occur as evidenced by the sharp break in slope
$dT_{c}/dP.$ At higher pressures $T_{c}(P)$ passes through a minimum before
dropping below 4 K at 67 GPa. Although it is possible that at or above 67 GPa
Li has transformed into the paired semiconducting/insulating state envisioned
by Neaton and Ashcroft, we have no clear evidence for this. At 67 GPa our Li
sample remains opaque to transmitted light in the visible region; from this
one can infer that Li is either still a metal or a semiconductor with a band
gap less than 1.7 eV. A more complete discussion of these results, including
evidence that Li is a type I superconductor, is given in Ref. \cite{deemyad1}.

Following the experiments on Li metal, two further alkali metals, Na and K,
were subjected to nearly hydrostatic dense He pressure in a search for
superconductivity. Shi \textit{et al.} \cite{shi1} have carried out an
electronic structure calculation on K and Rb and predict an onset of
superconducting near 13 GPa for K and 8 GPa for Rb with a rapid rise in
$T_{c}$ with pressure, in analogy with Li. In our experiment on Na metal to 65
GPa, however, no superconducting transition could be detected above 4 K. K
metal also failed to become superconducting above 4 K to 43.5 GPa pressure. In
a second experiment on K a nonsuperconducting MoW gasket was used which
permitted measurements to lower temperatures. In this experiment no
superconductivity was detected in K above 1.5 K to 35 GPa pressure. In a
quasihydrostatic experiment to 21 GPa by Ullrich \textit{et al. }%
\cite{ullrich}, no superconductivity was found in Rb above 0.05 K .

It would seem almost certain that all alkali metals will become
superconducting under sufficient pressure; after all, both the lightest (Li)
and heaviest (Cs) alkali metals do become superconducting! Further experiments
are underway over expanded temperature/pressure ranges to search for
superconductivity in Na, K, and Rb.

\bigskip

\noindent\textbf{Acknowledgments.} \ The authors are grateful to N.W. Ashcroft
and J.B. Neaton for stimulating discussions. Research is supported by NSF
Grant No. DMR-0404505.

\end{document}